\documentstyle[12pt]{article}

\begin{document}
\pagestyle{empty}
\begin{flushright}
{CERN-TH/95-204 \\
 hep-lat/9507024}
\end{flushright}
\vspace*{5mm}
\begin{center}
{\bf ELECTROWEAK PHASE TRANSITION AND} \\[0.3cm]
{\bf NUMERICAL SIMULATIONS IN THE SU(2) HIGGS MODEL}       \\
\vspace*{1.5cm}
{\bf I. Montvay$^1$}                             \\
\vspace{0.3cm}
Theoretical Physics Division, CERN               \\
CH-1211 Geneva 23, Switzerland                   \\
and                                              \\
Deutsches Elektronen-Synchrotron DESY,           \\
Notkestr.\,85, D-22603 Hamburg, Germany          \\
\vspace*{2cm}
{\bf ABSTRACT}                                   \\ \end{center}
\vspace*{0.3cm}
\noindent

 Recent progress in non-perturbative investigations of the electroweak
 phase transition is reviewed, with special emphasis on numerical
 simulations in the four-dimen\-sional SU(2) Higgs model.
\vspace*{2cm}

\begin{center}
{\em Lecture given at the $3^e$ Colloque Cosmologie,        \\
Paris, June 1995.}
\end{center}
\vspace*{1cm}

\begin{flushleft}
CERN-TH/95-204 \\
July 1995

$^1$ e-mail address: montvay@surya20.cern.ch
\end{flushleft}
\vfill\eject

\newcommand{\be}{\begin{equation}}
\newcommand{\ee}{\end{equation}}
\newcommand{\half}{\frac{1}{2}}
\newcommand{\rar}{\rightarrow}
\newcommand{\lar}{\leftarrow}
\newcommand{\LCB}{\raisebox{-0.3ex}{\mbox{\LARGE$\left\{\right.$}}}
\newcommand{\RCB}{\raisebox{-0.3ex}{\mbox{\LARGE$\left.\right\}$}}}

\setcounter{page}{1}
\pagestyle{plain}

\begin{center}
{\bf ELECTROWEAK PHASE TRANSITION AND} \\
{\bf NUMERICAL SIMULATIONS IN THE SU(2) HIGGS MODEL}       \\
\vspace*{1.5cm}
{I. MONTVAY}                                     \\
\vspace{0.3cm}
Theoretical Physics Division, CERN               \\
CH-1211 Geneva 23, Switzerland                   \\
and                                              \\
Deutsches Elektronen-Synchrotron DESY,           \\
Notkestr.\,85, D-22603 Hamburg, Germany          \\
\vspace*{1cm}
{ABSTRACT}                                       \\
\vspace*{0.3cm}
\parbox{13cm}{{\small
\hspace*{1em}
 Recent progress in non-perturbative investigations of the electroweak
 phase transition is reviewed, with special emphasis on numerical
 simulations in the four-dimensional SU(2) Higgs model.}}
\end{center}

\vspace*{0.5cm}
\begin{flushleft}
{\bf 1. Introduction}
\end{flushleft}

 The symmetry restoration in the electroweak Standard Model at high
 temper\-ature$^{\ref{KIRLIN}}$ is an interesting feature in the history
 of the early Universe.
 The anomalous baryon number violating processes are fast for
 temperatures above the electroweak phase transition of symmetry
 restoration, and are getting very slow soon after it.
 It is also possible that the baryonic excess was produced during the
 phase transition itself$^{\ref{KURUSH}}$.
 The conditions for this are: a strong enough first order phase
 transition and a sufficiently strong CP-violation in the Standard
 Model.
 Concerning the strength of CP-violation, in spite of some
 optimism$^{\ref{FARSHA}}$, serious doubts have been
 raised$^{\ref{GAORPE}}$.
 In order to determine the properties of the electroweak phase
 transition one needs non-perturbative studies, which can be performed
 within the framework of lattice regularization.
 In this lecture a short review will be given about some recent
 developments in this field.
 The discussion will be centred around a recent series of numerical
 simulations performed in the four-dimensional SU(2) Higgs
 model$^{\ref{PHYSLETT}-\ref{CSFOHEHE}}$.

\vspace*{0.3cm}
\begin{flushleft}
{\it 1.1. Dimensional reduction}
\end{flushleft}

 The main difficulty in numerical simulations of the thermodynamics
 of the electroweak plasma on the lattice has been identified in
 earlier studies$^{\ref{BUILKRSCH}}$: as a consequence of the weak
 gauge coupling, the observed first order phase transition gets
 very weak with increasing Higgs boson mass ($M_H$) already near
 $M_H \simeq M_W$ (with $M_W$ the $W$-boson mass).
 The screening masses (inverse correlation lengths) near the phase
 transition become much smaller than the temperature.
 It becomes very difficult to deal with this two-scale problem in a
 numerical simulation.

 This suggests a reduction of the problem to a three-dimensional field
 theory by integrating out the heavy modes with non-zero Matsubara
 frequencies$^{\ref{APPPIS}-\ref{LANDSMAN}}$.
 The simplest possibility is to consider the three-dimensional
 fundamental Higgs model with couplings determined by a perturbative
 matching procedure$^{\ref{FAKARUSH}}$.
 This amounts to choosing some perturbatively well calculable set of
 parameters of the four-dimensional continuum theory (usually derived
 from the effective potential in the phase with broken symmetry)
 and setting them equal to the corresponding parameters in the
 reduced three-dimensional continuum theory.
 Since the three-dimensional model is super-renormalizable,
 the correspondence of the renormalized parameters to the bare
 parameters in lattice regularization can be determined, near the
 continuum limit, in a two-loop perturbative
 calculation$^{\ref{LAINE}}$.
 In this way the non-perturbative properties of the phase transition
 can be numerically studied in lattice
 simulations$^{\ref{FAKARUSH},\ref{ILKRPESCH}}$, which are less
 demanding than the direct lattice simulations in the original
 four-dimensional theory.

 Originally, the three-dimensional reduced model contains also a
 Higgs field in the adjoint (triplet) representation, but its mass
 is governed by chromo-electric screening, and hence it is heavier
 than the Higgs doublet field and the three-dimensional gauge field,
 which gets its mass from chromo-magnetic screening.
 Therefore, at least in a first approximation, one can also neglect
 the adjoint Higgs field and stay with a simple three-dimensional
 fundamental SU(2) Higgs model.
 Further reduction, leading to scalar degrees of freedom alone, seems
 to give qualitatively different results$^{\ref{KANEPA}}$, and hence
 is not advisable.

 The obvious question about reduction is the effect of the necessary
 approximations on the non-perturbative results.
 In order to discuss this, it is advantageous to think about the
 four-dimensional continuum theory as the continuum limit of a
 lattice-regularized model.
 Although the continuum limit of the SU(2) Higgs model (and also of the
 electroweak sector of the Standard Model) is believed to be trivial,
 the renormalized couplings go to zero only logarithmically with the
 lattice spacing.
 Therefore the continuum theory with non-zero renormalized couplings
 can be realized to a very good approximation along the {\em lines of
 constant physics} defined by the constant values of two dimensionless
 parameters: the mass ratio $R_{HW} \equiv M_H/M_W$ and the
 renormalized gauge coupling $g_R^2$ defined, for instance, from the
 screened (Yukawa) potential of external charges$^{\ref{MONMUN}}$.
 The points of the lines of constant physics can be parametrized by
 the logarithm of the inverse $W$-boson mass in lattice units:
 $\tau \equiv \log(aM_W)^{-1}$.
 Due to the triviality of the continuum limit, every line of constant
 physics with non-zero couplings ends on the boundary of the bare
 parameter space at a finite value of $\tau=\tau_{\rm max}$, but if
 the Higgs boson mass is not close to the triviality upper
 limit$^{\ref{LANMON}}$ ($M_H \simeq 9M_W$), the value of
 $\tau_{\rm max}$ can be very large.
 In such cases, since lattice artefacts go to zero by powers of
 the lattice spacing, the continuum theory is very well approximated.

 In a given point of a line of constant physics, with large $\tau$,
 the heavy and high-momentum modes can be integrated out by
 considering the effective action of averaged
 fields$^{\ref{WETEWE},\ref{PAPEPO}}$
 or by applying block-spin transformations$^{\ref{KEMAPA}}$.
 In fact, for the purpose of obtaining a three-dimensional lattice
 model it is simplest to take the block-spin approach and to perform
 so many block-spin transformations that in the imaginary
 time (or inverse temperature) direction only a single layer of
 points is left over.
 In this way a three-dimensional lattice action is obtained for the
 light and low-momentum modes, which is exactly equivalent to the
 original four-dimensional lattice action.
 (Here we assume the ideal situation of a {\em perfect action}, when
 no approximations are made in the blocking procedure.)

 The three-dimensional lattice action contains, in general, many
 couplings.
 Besides the simplest nearest neighbour couplings, there are also less
 local terms, with non-locality at least as large as the inverse
 temperature represented by the periodicity in the fourth dimension
 of the original four-dimensional lattice.
 How well this multi-parameter action can be approximated by the
 simplest local action corresponding to the discretization of the
 simplest local three-dimensional continuum theory is an open question.
 The connection between the multi-parameter action and the simple
 local action giving the best approximation will, in general, also
 depend on the lattice spacing.
 This can be seen, for instance, if the integration of heavy degrees of
 freedom is done in two-loop perturbation theory with momentum
 cut-off$^{\ref{JAKOVAC}}$.
 In fact, the continuum limit in case of the three-dimensional
 block-spin action is not the same as the continuum limit for the
 simplest local three-dimensional action.
 Starting from different points of some four-dimensional line of
 constant physics with increasing scale parameter $\tau$, the
 lattice spacing for the three-dimensional block-spin action
 is not changing at all.
 (The couplings in the action are, of course, changing.)
 One can also say that one should not take the continuum limit of an
 effective field theory beyond the scale where ``new physics'' appears.

 The danger in taking only the simplest local three-dimensional action
 is that the approximation given by it may be of different quality
 in different parts of the parameter space: the action optimized in
 the perturbative region, deeply inside the Higgs phase, can deviate
 more from the block-spin action near the phase transition and in the
 symmetric phase.
 In fact, it is plausible that in the symmetric phase, which is
 similar to the high-temperature phase of a non-Abelian pure gauge
 theory, the r\^ole of non-locality is more important.

 This shows that the numerical simulations of the electroweak phase
 transition in its unreduced four-dimensional form are important,
 even if finally the dimensional reduction, in some form, will turn
 out to be a good approximation.

\vspace*{0.3cm}
\begin{flushleft}
{\it 1.2. Lattice action}
\end{flushleft}

 The numerical simulations in the four-dimensional finite temperature
 theory have been restricted up to now to the SU(2) Higgs model.
 This is a good and sufficiently simple theoretical laboratory,
 where the qualitative features of the symmetry restoring phase
 transition can be understood.
 The well known continuum action of this model is given in terms of
 the SU(2) gauge field $W_\mu^r(x)\; (\mu=1,2,3,4;\; r=1,2,3)$
 and the complex scalar doublet field
 $\phi_\alpha(x)\; (\alpha=1,2)$.
 The coupling parameters are: the (squared) gauge coupling $g^2$ and
 the quartic scalar self-coupling $\lambda$.
 In addition, there is a single mass parameter belonging to the
 scalar field, which can be represented in the Higgs phase by the
 vacuum expectation value $v$.

 The Euclidean lattice action of the SU(2) Higgs model can be written
 as
$$
S[U,\varphi] = \beta \sum_{pl}
\left( 1 - \frac{1}{2} {\rm Tr\,} U_{pl} \right)
$$
\be \label{eq01}
+ \sum_x \left\{ \half{\rm Tr\,}(\varphi_x^+\varphi_x) +
\lambda \left[ \half{\rm Tr\,}(\varphi_x^+\varphi_x) - 1 \right]^2 -
\kappa\sum_{\mu=1}^4
{\rm Tr\,}(\varphi^+_{x+\hat{\mu}}U_{x\mu}\varphi_x)
\right\} \ .
\ee
 Here $U_{x\mu}$ denotes the SU(2) gauge link variable, $U_{pl}$
 is the product of four $U$'s around a plaquette, and
 $\varphi_x$ is a complex $2 \otimes 2$ matrix in isospin space
 describing the Higgs scalar field.
 The bare parameters in the action are $\beta \equiv 4/g^2$ for
 the gauge coupling, $\lambda$ for the scalar quartic coupling and
 $\kappa$ for the scalar hopping parameter related to the bare
 mass square $\mu_0^2$ by $\mu_0^2 = (1-2\lambda)\kappa^{-1} - 8$.
 In what follows we set the lattice spacing to 1 ($a=1$),
 therefore all the masses and correlation lengths, etc., will always be
 given in lattice units, unless otherwise stated.

 The large scale numerical
 simulations$^{\ref{PHYSLETT}-\ref{CSFOHEHE}}$, which will be discussed
 in the present lecture, have been performed mainly on the APE-Quadrics
 massive parallel computers at DESY, and in a smaller part also on the
 CRAY's of HLRZ-J\"ulich.
 The published results refer mainly to two Higgs boson masses:
 $M_H \simeq 20\;{\rm GeV}$ and $\simeq 50\;{\rm GeV}$,
 always assuming that
 $M_W \equiv 80\;{\rm GeV}$.
 Recently, an intermediate Higgs boson mass $M_H \simeq 35\;{\rm GeV}$
 has also been considered$^{\ref{CSFOHEHE}}$, mainly in order to be
 able to compare the results with past and future results of numerical
 simulations in the reduced three-dimensional models.
 This work is still going on and will be published soon$^{\ref{SOON}}$.

 In the optimized updating procedure for the creation of the sample
 of field configurations, besides the algorithms described in detail
 previously$^{\ref{NUCLPHYS}}$, also scalar field
 reflections$^{\ref{BUNK}}$ have been applied.
 In fact, this latter algorithm turned out to be very efficient in
 reducing the autocorrelations among field configurations, hence in
 reducing the necessary computer resources.

\vspace*{0.5cm}
\begin{flushleft}
{\bf 2. Latent heat}
\end{flushleft}

 One of the most important physical quantities characterizing first
 order phase transitions is the latent heat, which is the discontinuity
 of the energy density $\Delta\epsilon$ across the phase transition.
 In the SU(2) Higgs model on the lattice its dimensionless ratio with
 the fourth power of the transition temperature $T_c$ can be
 obtained$^{\ref{PHYSLETT}}$ from
\be \label{eq02}
\frac{\Delta\epsilon}{T_c^4} = L_t^4
\left\langle \frac{\partial\kappa}{\partial\tau}
\cdot 8\Delta L_{\varphi,x\mu}
- \frac{\partial\lambda}{\partial\tau} \cdot \Delta Q_x
- \frac{\partial\beta}{\partial\tau}
\cdot 6\Delta P_{pl} \right\rangle \ .
\ee
 Here $L_t$ is the lattice extension in the inverse temperature
 direction, the partial derivatives of the bare couplings are taken
 along lines of constant physics, and
$$
L_{\varphi,x\mu} \equiv
\half {\rm Tr\,}(\varphi^+_{x+\hat{\mu}}U_{x\mu}\varphi_x) \ ,
$$
$$
Q_x \equiv \left[ \half{\rm Tr\,}(\varphi^+_x \varphi_x)
- 1 \right]^2 \ ,
$$
\be \label{eq03}
P_{pl} \equiv 1 - \half {\rm Tr\,} U_{pl} \ .
\ee
 The signs of the discontinuities such as $\Delta\epsilon$ etc.\ are
 usually defined as differences of values in the symmetric phase minus
 the Higgs phase.

 For the numerical evaluation of the latent heat the knowledge of the
 lines of constant physics is required.
 Since in the Standard Model, for not very high Higgs boson masses,
 we are interested in weak couplings, estimates of $\lambda(\tau)$
 and $\beta(\tau)$ can be obtained from the perturbative
 renormalization group equations.
 Similarly, $\kappa(\tau)$ can be estimated using one-loop lattice
 effective potentials$^{\ref{PHYSLETT},\ref{NUCLPHYS}}$.
 These estimates can then be ckecked in numerical simulations
 by determining $R_{HW}$ and $g_R^2$ numerically.
 If necessary, one can also further tune the bare parameters in order to
 reach constant values of $R_{HW}$ and $g_R^2$.
 In practice, in the investigated parameter range, the perturbative
 estimates are usually sufficient.
 Assuming that the temperature is equal to $T_c$, the scale
 changes are determined by the lattice extensions in the fourth
 direction: $T_c=1/L_t$.
 Taking $L_t=2,3,4,\ldots$, we get for the changes in the scale
 parameter $\Delta\tau=\log(3/2),\log(4/2),\ldots$.

 The other ingredients in Eq.~\ref{eq02} are the discontinuities
 of the global quantities defined in Eq.~\ref{eq03}.
 In finite spatial volumes the discontinuities are round-off by
 inverse volume ($(L_xL_yL_z)^{-1}$) effects.
 These finite volume effects can be eliminated, in principle, by
 a cumbersome extrapolation to infinite volume.
 Another possibility is to use the metastability of the two phases
 in the vicinity of the phase transition.
 This implies that, in large enough volumes, during the Monte Carlo
 updating the field configuration remains in one metastable phase
 for a very long time.
 In this way one is able to determine expectation values characterizing
 single phases, and hence e.~g.\ the required discontinuities.
 One has to have in mind, however, that in this way an independent
 precise determination of the location of the phase transition is
 necessary, because the metastability extends over a finite range of
 parameters.
 The uncertainty of the phase transition parameters appears in the
 errors of the discontinuities$^{\ref{SOON}}$.
\begin{figure}
\vspace{8.5in}
\includegraphics{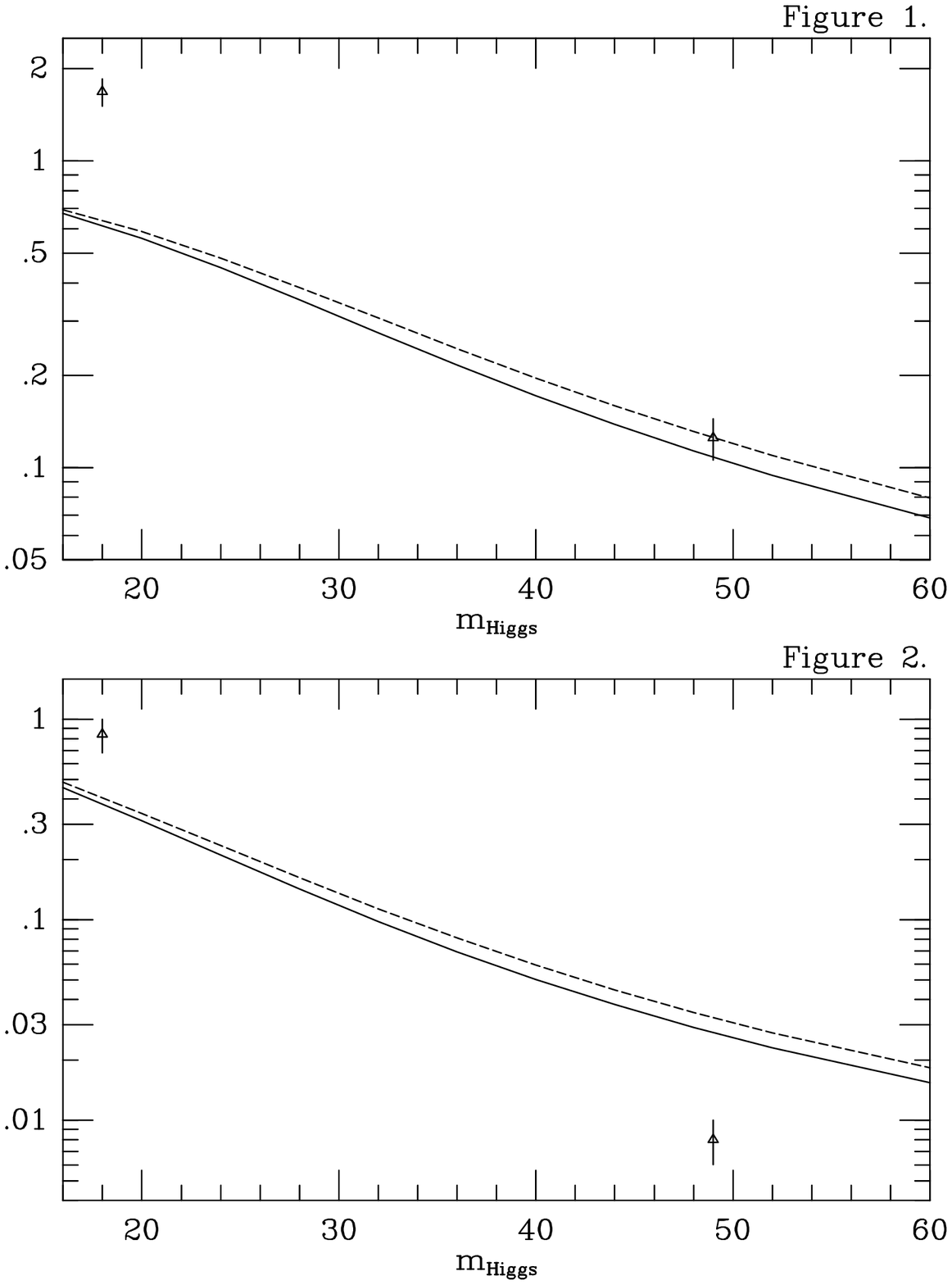}
\end{figure}

 First results$^{\ref{PHYSLETT},\ref{NUCLPHYS}}$ on
 $\Delta\epsilon/T_c^4$ as a function of the Higgs boson mass
 are shown in Fig.~1, together with a comparison
 to two-loop resummed perturbation theory$^{\ref{BUFOHE}}$.
 As can be seen, the results of the numerical simulation (shown by
 triangles) agree qualitatively well with perturbation theory.
 The latent heat characterizing the strength of the first order phase
 transition is decreasing considerably between
 $M_H \simeq 20\; {\rm GeV}$ and $M_H \simeq 50\; {\rm GeV}$.

\vspace*{0.5cm}
\begin{flushleft}
{\bf 3. Interface tension}
\end{flushleft}

 Besides the latent heat, another important physical quantity
 characterizing the first order  phase transitions is the interface
 tension ($\sigma$): the free energy per unit area of the wall
 separating the two phases.
 Many characteristics of the nucleation dynamics are determined by
 these two quantities.
 The radius of the critical droplets is proportional to
 $\sigma/\Delta\epsilon$.
 The nucleation rate is approximately determined by
\be \label{eq04}
R_n \equiv \frac{\sigma^3}{(\Delta\epsilon)^2 T_c}
= \frac{(\sigma/T_c)^3}{(\Delta\epsilon/T_c^4)^2} \ .
\ee
 The typical droplet distance is roughly proportional to $R_n^\half$.
 In fact, the quantity $R_n$ defined in Eq.~\ref{eq04} can be taken
 as a good characterization of the strength of the first order phase
 transition: for $R_n = {\cal O}(1)$ we have a strong transition,
 whereas $R_n \ll 1$ corresponds to a weak one.

 In numerical simulations one can exploit the tunnelling between the
 two nearly equal minima of the effective potential, which correspond
 to the two phases.
 On large enough lattices one can study the properties of the mixed
 phase when, as a result of tunnelling, both phases are simultaneously
 present in the system.
 For this purpose an elongated geometry of the lattice is advantageous.
 In a four-dimensional simulation, when the fourth dimension gives
 the inverse temperature ($T=L_t^{-1}$), one can take an extension
 $L_z$ much larger than the other two ``transverse'' directions:
 $L_z \gg L_x=L_y$.
 In this case, for large enough transverse extensions when the mixed
 phase appears, there are transverse walls separating the two phases.
 For periodic boundary conditions, and not too large $L_z$, there are
 just two walls and two regions (one for each phase).

 This situation can be used to determine the interface tension in
 several ways.
 For instance, one can determine $\sigma$ from the small energy
 difference in the transfer matrix between the asymmetric and
 symmetric combinations of the two phases$^{\ref{JJMMTW}}$.
 Another possibility is to measure the doubly peaked distribution of
 some order parameter, when $\sigma$ is given by the ratio of the
 maxima to the flat region between the two peaks.
 This can be effectively done by the ``multicanonical'' simulation
 method$^{\ref{BERNEU}}$.
 In the ``two-coupling'' method$^{\ref{POTREB}}$ one divides the
 long lattice extension into two halves with different bare parameters:
 in one half slightly above ($\kappa > \kappa_c$) in the other half
 slightly below ($\kappa < \kappa_c$) of the phase transition.
 In the range $20\; {\rm GeV} < M_H < 50\; {\rm GeV}$ all three
 methods are applicable and give compatible
 results$^{\ref{NUCLPHYS},\ref{CSFOHEHE}}$.
 In Fig.~2 the obtained interface tension at
 $M_H \simeq 20\; {\rm GeV}$ and $\simeq 50\; {\rm GeV}$ is shown
 and compared with two-loop resummed perturbation
 theory$^{\ref{BUFOHE}}$.
 At $M_H \simeq 35\; {\rm GeV}$ a recent result is$^{\ref{CSFOHEHE}}$
\be \label{eq05}
\left. \frac{\sigma}{T_c^3} \right|_{\rm 35\; GeV} = 0.06 \pm 0.01 \ .
\ee
 This shows that also the interface tension is a steeply decreasing
 function of the Higgs boson mass.

 The combination introduced in Eq.~\ref{eq04}, which is a good
 characterization for the strength of the phase transition, is
 decreasing from $R_n \simeq 0.23$ at $M_H \simeq 20\; {\rm GeV}$ to
 $R_n \simeq 0.00003$ at $M_H \simeq 50\; {\rm GeV}$.
 That is, the phase transition at $M_H \simeq 50\; {\rm GeV}$ becomes
 very weak indeed.

 The interface tension, more precisely $\sigma/T_c$, plays a decisive
 r\^ole also in determining the minimal spatial lattice size required
 for an accurate numerical investigation of first order phase
 transitions.
 In order to determine discontinuities of global averages, screening
 lengths in separated metastable phases and other similar
 characteristics of the transition, one has to reach a mixed phase
 situation, which can be characterized by a well developed double
 peak structure of the order parameter distributions.
 The double peak structure appears when
 $L_xL_y\sigma/T_c = {\cal O}(1)$, with $L_x,L_y \leq L_z$ the
 transverse lattice extensions.
 The lattice in this volume should be fine enough to also resolve
 the interface structure, which becomes thiner for increasing $M_H$.
 These are physical requirements setting a minimum on the spatial
 lattice size, which are practically more important than the conditions
 imposed by the smallness of ${\cal O}(a)$ lattice artefacts.
 This implies that the requirements on spatial lattice sizes are
 essentially the same in dimensionally reduced three-dimensional
 simulations as in unreduced four-dimensional ones.
 The gain in computational resources by using the simplest local
 three-dimensional action is essentially equal to a factor $L_t$.
 The consequence is that, for instance, the extension of the present
 numerical studies to Higgs boson masses beyond
 $M_H \simeq 50\; {\rm GeV}$ is not much easier in three dimensions
 than in four.

\vspace*{0.5cm}
\begin{flushleft}
{\bf 4. Discussion}
\end{flushleft}

 The large scale numerical simulations discussed in the previous
 sections show that the symmetry restoring electroweak phase transition
 is of first order, with fast decreasing strength, in the Higgs boson
 mass range $20\; {\rm GeV} < M_H < 50\; {\rm GeV}$.
 This conclusion is in qualitative agreement with two-loop resummed
 perturbation theory$^{\ref{BUFOHE}}$.
 At $M_H \simeq 50\; {\rm GeV}$ the strength of the phase transition
 is definitely not enough to produce a strong enough out-of-equilibrium
 situation in the early Universe for the creation of the observed
 baryon asymmetry.
 The observed difficulty of identifying the first order phase transition
 near $M_H \simeq 80\; {\rm GeV}$ shows$^{\ref{BUILKRSCH}}$ that in the
 range $50\; {\rm GeV} < M_H < 100\; {\rm GeV}$ presumably no surprises
 are to be expected.
 Nevertheless, before a definite conclusion, the non-perturbative studies
 have to be extended towards higher Higgs boson masses, which are
 actually the experimentally allowed ones in the minimal Standard Model.
 This seems feasible in both unreduced and dimensionally reduced SU(2)
 Higgs models.

\vspace*{0.5cm}
\begin{flushleft}
{\bf 5. Acknowledgements}
\end{flushleft}

 I thank Zolt\'an Fodor and Andr\'as Patk\'os for stimulating
 discussions.
 The kind hospitality of Norma S\'anchez and Hector de Vega during this
 nice Colloquium is gratefully acknowledged.

\vspace*{0.5cm}
\begin{flushleft}
{\bf 6. References}
\end{flushleft}
\begin{enumerate}
\item\label{KIRLIN}
D.A. Kirzhnitz,
{\it JETP Lett.} {\bf 15} (1972) 529; \\
D.A. Kirzhnitz and A.D. Linde,
{\it Phys. Lett.} {\bf B72} (1972) 471;
{\it Ann. Phys.} {\bf 101} (1976) 195.
\item\label{KURUSH}
V.A. Kuzmin, V.A. Rubakov and M.E. Shaposhnikov,
{\it Phys. Lett.} {\bf B155} (1985) 36.
\item\label{FARSHA}
G.R. Farrar and M.E. Shaposhnikov,
{\it Phys. Rev. Lett.} {\bf 70} (1993) 2833.
\item\label{GAORPE}
M.B. Gavela, M. Lozano, J. Orloff and O. P\`ene,
{\it Nucl. Phys.} {\bf B430} (1994) 345;   \\
M.B. Gavela, P. Hernandez, J. Orloff, O. P\`ene and C. Quimbay,
{\it Nucl. Phys.} {\bf B430} (1994) 382.
\item\label{PHYSLETT}
F. Csikor, Z. Fodor, J. Hein, K. Jansen, A. Jaster and I. Montvay,
{\it Phys. Lett.} {\bf B334} (1994) 405.
\item\label{NUCLPHYS}
Z. Fodor, J. Hein, K. Jansen, A. Jaster and I. Montvay,
{\it Nucl. Phys.} {\bf B439} (1995) 147.
\item\label{BIELEFELD}
F. Csikor, Z. Fodor, J. Hein, K. Jansen, A. Jaster and I. Montvay,
{\it Nucl. Phys. B (Proc. Suppl.)} {\bf 42} (1995) 569.
\item\label{CSFOHEHE}
F. Csikor, Z. Fodor, J. Hein and J. Heitger,
CERN preprint, 1995  \\   ({hep-lat/9506029}).
\item\label{BUILKRSCH}
B. Bunk, E.M. Ilgenfritz, J. Kripfganz and A. Schiller,
{\it Phys. Lett.} {\bf B284} (1992) 371;
{\it Nucl. Phys.} {\bf B403} (1993) 453.
\item\label{APPPIS}
T. Appelquist and R. Pisarski,
{\it Phys. Rev.} {\bf D23} (1981) 2305.
\item\label{LANDSMAN}
N.P. Landsman,
{\it Nucl. Phys.} {\bf B322} (1989) 489.
\item\label{FAKARUSH}
K. Farakos, K. Kajantie, K. Rummukainen and M. Shaposhnikov,
{\it Nucl. Phys.} {\bf B407} (1993) 356;
{\bf B425} (1994) 67;
{\bf B442} (1995) 317.
\item\label{LAINE}
M. Laine,
Helsinki preprint, 1995 ({hep-lat/9504001}).
\item\label{ILKRPESCH}
E.M. Ilgenfritz, J. Kripfganz, H. Perlt and A. Schiller,
Heidelberg preprint, 1995 ({hep-lat/9506023}).
\item\label{KANEPA}
F. Karsch, T. Neuhaus and A. Patk\'os,
{\it Nucl. Phys.} {\bf B441} (1995) 629.
\item\label{MONMUN}
I. Montvay and G. M\"unster,
{\em Quantum Fields on a Lattice} (Cambridge University Press, 1994).
\item\label{LANMON}
W. Langguth and I. Montvay,
{\it Z. Phys.} {\bf C36} (1987) 725.
\item\label{WETEWE}
C. Wetterich,
{\it Nucl. Phys.} {\bf B352} (1991) 529;  \\
N. Tetradis and C. Wetterich,
{\it Nucl. Phys.} {\bf B398} (1993) 659.
\item\label{PAPEPO}
A. Patk\'os, P. Petreczky and J. Polonyi,
Budapest preprint, 1995  \\   ({hep-ph/9505221}).
\item\label{KEMAPA}
U. Kerres, G. Mack and G. Palma,
DESY preprint, 1995 ({hep-lat/9505008}).
\item\label{JAKOVAC}
A. Jakov\'ac,
Budapest preprint, 1995 ({hep-ph/9502313}).
\item\label{SOON}
Z. Fodor, A. Jaster, J. Hein and I. Montvay,
in preparation.
\item\label{BUNK}
B. Bunk,
{\it Nucl. Phys. B (Proc. Suppl.)} {\bf 42} (1995) 566.
\item\label{BUFOHE}
Z. Fodor and A. Hebecker,
{\it Nucl. Phys.} {\bf B432} (1994) 127;   \\
W. Buchm\"uller, Z. Fodor and A. Hebecker,
DESY preprint, 1995  \\   ({hep-ph/9502321}).
\item\label{JJMMTW}
K. Jansen, J. Jers\'ak, I. Montvay, G. M\"unster, T. Trappenberg
and U. Wolff,
{\it Phys. Lett.} {\bf B213} (1988) 203.
\item\label{BERNEU}
B. Berg and T. Neuhaus,
{\it Phys. Rev. Lett.} {\bf 68} (1992) 9.
\item\label{POTREB}
J. Potvin and C. Rebbi,
{\it Phys. Rev. Lett.} {\bf 62} (1989) 3062.
\end{enumerate}

\end{document}